\documentclass[aps,prX,twocolumn,showpacs]{revtex4}  
\usepackage{graphicx}  % needed for figures
\usepackage{bm}        % for math
\usepackage{amssymb}   % for math
\usepackage{verbatim}   
\usepackage{subfigure}

\begin{document}
\title{Multichannel Interference In High-Order Harmonic Generation
From Ne$^{+}$ Driven By Ultrashort Intense Laser Pulse}
\author{O. Hassouneh, A. C. Brown and H. W. van der Hart}

\affiliation{Centre for Theoretical Atomic, Molecular and Optical Physics,
School of Mathematics and Physics,
Queen's University Belfast, Belfast BT7 1NN, United Kingdom}

\date{\today}

\begin{abstract}
We apply the time-dependent R-matrix method to investigate harmonic
generation from Ne$^{+}$ at
a wavelength of 390 nm and intensities up to $10^{15}$  Wcm$^{-2}$. The
$1s^{2}2s^{2}2p^{4}$ $(^{3}P^{e}, ^{1}D^{e},$ and $^{1}S^{e})$ states of Ne$^{2+}$
are included as residual-ion states to assess the influence
of interference between photoionization channels associated with these thresholds.
The harmonic spectrum is well approximated by calculations in which only the
$^3P^e$ and $^1D^e$ thresholds are taken into account, but no satisfactory
spectrum is obtained when a single threshold is taken into account.
Within the harmonic plateau, extending to about 100 eV, individual harmonics can be
suppressed at particular intensities when all Ne$^{2+}$ thresholds are taken
into account. The suppression is not observed when only a single threshold
is accounted for. Since the suppression is dependent on intensity, it may be difficult to observe experimentally.  
\end{abstract}
\maketitle

\section{Introduction}
High-order harmonic generation (HG) has been investigated 
intensively since its observation more than three decades ago \cite{Ago}. It has remained a process of great interest
due to its potential as a source of XUV radiation \cite{Zhao}. Since the light generated by HG is coherent, HG 
is the key for the production of ultrashort pulses of light attosecond duration \cite{Paul, Hen}.
These ultrashort pulses
have been applied to a variety of time-resolved studies, enabling
researchers to explore ultrafast electron dynamics occurring on the sub-femtosecond
time scale \cite{Sch, Kl}. Examples of these explorations include the investigation of dynamics in
inner shell vacancies for Kr atoms \cite{Dre}, the time-resolved investigation of
laser induced tunnel ionization \cite{Uib} and the study of electron dynamics during
photoemission from a tungsten surface \cite{Cav}. 

Harmonic generation is also of interest as a measurement tool in the study of molecular dynamics and structure.
Different harmonics are generated at slightly different times, and this characteristic has been
used to demonstrate differences in the period of molecular vibrations \cite{S}. 
More recently, a harmonic spectroscopy technique allows photochemical reactions to be 
measured in real time \cite{Wor1}; this method has the ability to monitor the
dynamic electron density as the electrons are transferred between atoms in a
molecule during chemical reactions. 
Furthermore, HG has been applied to investigate how 
Br atoms move apart in the dissociation of Br$_{2}$ molecules \cite{Wor2}.

HG is generally understood in terms of the quasi-classical,
three-step model \cite{Cor1993}. Three different stages of electron dynamics need to 
be included: first, the ejection of an electron from the atom, second, the motion of this
ejected electron in the laser field and, third, its photo-recombination with the parent ion.
The three-step model has proven to be very useful for understanding many of the features
of HG. It provides intuitive understanding about the highest harmonic that can be 
observed and the timing delays between different harmonics.

One of the key findings in recent experiments is that HG in molecules is significantly
affected by interferences between ionization channels associated with different ionization
thresholds \cite{Shi}. Although the three step model is suitable for noble-gas targets which
have an isolated lowest ionization threshold, it may be less suitable for systems
with several, closely spaced low-lying ionization thresholds. For these systems,
the electronic dynamics may become
quite complicated, leading to amplitude and phase differences between different
ionization pathways, phase differences between electron trajectories in the laser
field associated with different ionization thresholds, and amplitude differences in the 
recombination step. Hence, although the underlying mechanism of HG
does not change for systems with multiple low-lying ionization thresholds, additional
physics merits consideration. Recent theoretical studies on HG in atoms have
demonstrated that interference effects between states associated with different
ionization thresholds can play a notable role in atomic systems as
well \cite{Pab1,Pab2,Bro1,Bro2,Sta}. 

Atomic systems present significant advantages in developing understanding of
HG in systems with multiple ionization thresholds. Many atomic systems
have multiple low-lying ionization thresholds. Atomic systems can
be described with very good accuracy, enabling a detailed investigation of the effect
electronic interactions have on the competition between different pathways. Accurate theoretical
methods are available: for example,
time-dependent R-matrix (TDRM) theory was developed at Queen's University Belfast to investigate the influence of
electron interactions on atomic dynamics in intense fields \cite{Lys, Lys1}. Pabst and Santra
\cite{Pab1, Pab2} have developed the TD-CIS approach, whereas Ngoko Djiokap and Starace \cite{Sta}
have investigated harmonic generation in the two-electron He atom by solving the
full-dimensional Schr\"{o}dinger equation.

The accuracy of the TDRM approach to HG processes was verified by the comparison of HG  
spectra for He with those obtained by the HELIUM code \cite{Smy98, Bro2}. It has since been applied to investigate the multielectron response for several systems: the effects of resonances on HG in Ar \cite{Bro1},
and the effect of multiple ionization thresholds on HG in Ar$^{+}$
\cite{Bro3,Bro4}. In these Ar$^{+}$ studies, only a few harmonics appeared for photon energies exceeding the Ar$^+$
binding energy, and it was therefore difficult to identify clear interference effects in the plateau region.
The Ar$^{+}$ studies were carried out at $4\times 10^{14}$ Wcm$^{-2}$. At higher intensities, high ionization
probabilities made the accurate determination of the hamonic spectrum impossible. The study of HG from ions is not only of theoretical interest. It has been suggested that the very highest harmonics seen in experiment are generated by ionised atoms rather than neutral atoms \cite{Ze, Gi, Wa}.

In the present study, we have chosen to investigate Ne$^{+}$. Due to its
higher ionization potential, the intensity can be increased significantly before ionization leads to a loss of
accuracy in the harmonic spectrum. Ne$^{+}$ is therefore a more suitable ion for investigating
interference effects between pathways associated with different threshold. The energy gap between the Ne$^{+}$ ionization thresholds is about twice as large as
the energy difference between Ar$^{+}$ thresholds. However, the energy difference remains comparable
to the photon energy, so the interaction between channels associated with different thresholds
should still be strong.

The organization of the paper is as follows. In the following section, we present briefly
the theoretical background, giving an overview of TDRM theory, and details on the computations.
In Sec III we present harmonic spectra including the three low-lying ionization thresholds,
the $2s{^2}2p{^4}$ $^{3}P^{e}$, $^{1}D^{e}$, $^{1}S^{e}$ states of Ne$^{2+}$, in the
calculations for several intensities. We then look in detail at the role of the different ionization thresholds on
HG by presenting harmonic spectra for calculations in which subsets of the ionization thresholds have been
included. Since the total magnetic quantum number significantly affects harmonic
generation \cite{Bro4}, results for $M=0$ are presented in section \ref{sec:m0} and those for
$M=1$ in section \ref{sec:m1}.
Finally, we will summarize our results and conclusions.  

\section{THEORY}
\subsection{TDRM Theory}

In this report, we employ TDRM theory to study HG in Ne$^{+}$. TDRM theory is a fully non-perturbative
{\it ab initio} theory, which has been developed to describe the interaction of an intense
ultrashort light pulse with general multielectron atoms and atomic ions. In the theory, it is assumed
that the light field can be treated classically in the dipole approximation, that it is linearly polarized
along the $\hat{z}$-axis and that it is spatially homogeneous. At present, relativistic effects are not taken into
account.

TDRM theory is based upon R-matrix theory, in which space is divided into two distinct regions:
an inner region and an outer region. In the inner region, all electrons are contained within a distance
$a_{r}$ of the nucleus. In the outer region, one electron has separated from the residual ion, and is
at a distance greater than $a_{r}$ from the nucleus, while the other electrons remain confined within
a distance $a_{r}$ of the nucleus. In the inner region, electron exchange and correlation effects
between all pairs of electrons are described in full. However, in the outer region, exchange interactions
between the ejected electron and the electrons remaining near the residual ion can be neglected.
Hence, in this region only the laser field and the long-range (multipole) potential of the residual ion
are included for the motion of the outer electron.

To obtain the wavefunction for the initial state of Ne$^+$, which is fully contained within the inner
R-matrix region at time $t=0$, we solve the time-independent field-free Schr\"{o}dinger equation. The
inner-region wavefunction is expanded in terms of an R-matrix basis $\psi_{k}(X_{N+1})$, 
given by \cite{Burbook}
\begin{eqnarray}\nonumber
\psi_k(\mathbf{X}_{N+1}) &=& \mathcal{A} \sum_{pj}{\phi_{p}(\mathbf{X}_{N};\hat{r}_{N+1})r^{-1}_{N+1} c_{pjk} u_{j}(r_{N+1})} 
\\ 
&&+\sum_j \chi_j(\mathbf{X}_{N+1})d_{jk}.
\label{eq:rmexp}
\end{eqnarray}
$\mathcal{A}$ is the antisymmetrization operator, $\phi_p$ are channel functions
in which residual Ne$^{2+}$ ion states
are coupled with the spin and angular
parameters of the outer electron. $\mathbf{X}_{N+1}$ = $\mathbf{x}_{1}, \mathbf{x}_2, \ldots, \mathbf{x}_{N+1}$, where $\mathbf{x}_{i} = \mathbf{r}_{i}\sigma_{i}$ are the space and spin coordinates of the $ith$ electron.
The functions $u_j(r_{N+1})$ form a continuum basis set for the radial
wavefunction of the outer electron
inside the inner region. Correlation functions $\chi_j$ are  $N+1$-electron basis functions which vanish
at the boundary. The residual-ion states $\phi_p$ and correlation functions $\chi_j$ are constructed from
Hartree Fock, Ne$^{2+}$ orbitals \cite{Roe}, and the functions $u_j$ are orthogonalised with respect to these input orbitals.
The coefficients $c_{pjk}$ and $d_{jk}$ are obtained through diagonalization of the field-free
Hamiltonian.

Once we have obtained the initial state, we solve the time-dependent Schr\"odinger equation for an atom
in a light field on a discrete time grid of step size $\Delta{t}$ using the Crank-Nicolson technique
described by Lysaght \textit{et al} \cite{Lys}. We obtain the wavefunction at time $t=t_{m+1}$ from
the solution at $t=t_m$, by rewriting the Schr\"odinger equation
using the unitary Cayley form of the time
evolution operator exp(-$itH(t)$). This gives, correct to ${O(\Delta{t}}^3)$,
 \begin{equation}
 [H(t_{m+1/2} )-E]\Psi(\mathbf{X}_{N+1},t_{m+1} )=\Theta(\mathbf{X}_{N+1},t_{m}),  
 \label{eq:tdse}
 \end{equation}               
where
\begin{equation}
\Theta(\mathbf{X}_{N+1},t_{m})=-[H(t_{m+1/2} )+E]\Psi(\mathbf{X}_{N+1},t_{m}).
 \end{equation}                                  
In Eqs. (2) and (3), $E =2i/\Delta{t}$, with $\Delta{t}=t_{m+1}- t_{m}$. $H(t_{m+1/2})$
is the time-dependent Hamiltonian at the midpoint time of $t_{m}$ and $t_{m+1}$. In this
time-dependent Hamiltonian, the laser field is described in the dipole-length
gauge.

In order to solve Eq. (\ref{eq:tdse}), we apply different approaches to the
inner region and the outer region \cite{Lys}. Within the inner region, we expand the
time-dependent wavefunction in terms of the field-free R-matrix basis:
\begin{equation}
\Psi(\mathbf{X}_{N+1},t_{m+1}) = \sum_{k}{\psi_{k}(\mathbf{X}_{N+1}) A_{k}(t_{m})},
\label{eq:innerexp}
\end{equation}
so that the time-dependence is contained entirely within the coefficients $A_k$.
However, the continuum functions $u_j$ in the R-matrix basis expansion (\ref{eq:rmexp})
are non-vanishing at the boundary. Hence the Hamiltonian $H(t_{m+1/2})$
is not Hermitian in the inner region due to surface terms arising from
the kinetic energy operator, $-\frac{1}{2}{\nabla_{i}}^{2}$.
We introduce the Bloch operator, $L$, to cancel these terms, such that $H{(t_{m+1/2})}+L $ is Hermitian in the internal region.
\begin{equation}
L = \frac{1}{2}\delta(r-a)\frac{d}{dr},
\label{eq:bloch}
\end{equation}
Using this result we can rewrite Eq. (\ref{eq:tdse}) in the internal region as:
\begin{equation}
\Psi = {(H + L - E)}^{-1} L\Psi + {(H + L - E)}^{-1}\Theta.
\label{eq:tdinner}
\end{equation}  

Similar to standard R-matrix theory \cite{Burbook}, we now need outer-region information to set
the boundary conditions for $\Psi$ to solve Eq. (\ref{eq:tdinner}) in the inner
region \cite{Lys}.
Hence, the inner and outer regions are linked to each other through the so-called R-matrix:
\begin{equation}
\mathbf{R}_{pp^{\prime}}(E)=\frac{1}{(2a_r)}\sum_{kk^{\prime}}
{\omega_{pk}\left(\frac{1}{(H+L)_{(kk^{\prime})}-E}\right)\omega_{p^{\prime}k^{\prime}}}.
\end{equation} 
Here, $p$ and $p^{\prime}$ indicate channel functions $\phi_{p}$, whereas $k$ and $k^{\prime}$
indicate field-free eigenfunctions $\psi_k$. $\omega_{pk}$ indicates the surface amplitude of the
field-free eigenfunctions at $a_r$ with respect to each channel function $\phi_p$. In the present computational scheme, the
R-matrix is obtained through solving a system of linear equations rather than a diagonalization of $H+ L$. 

The wavefunctions in the inner and the outer regions are then connected to
each other at the boundary according to \cite{Burbook, Lys} by
\begin{equation}
\mathbf{F}(a_{r} )=\mathbf{R}a_{r}\mathbf{\bar{F}}(a_{r} )+ \mathbf{T}(a_{r}), 
\label{eq:frtrelation}
\end{equation}  
where the vector $\mathbf{F}$ is the reduced radial wavefunction of the
scattered electron, and $\mathbf{\bar{F}}$ its first derivative. Compared to standard R-matrix theory, an additional
inhomogenous term appears on the right hand side, which arises from the $\Theta$-term in Eq. (\ref{eq:tdse}).
This so-called T-vector is given by:
\begin{equation}
\mathbf T_{p}(a_{r})=\sum_{kk^{\prime}}{\omega_{pk}
\left({\frac{1}{(H+L)_{(kk^{\prime})}-E}}\right) \langle\psi_{k}(\mathbf{X}_{N+1})\vert\Theta \rangle}.
\end{equation}
The T-vector is determined together with the R-matrix in
the linear solver step. With this equation, we can determine the full wavefunction
in the inner region once we know the vector $F(a_r)$. In standard R-matrix theory,
this is achieved through setting boundary conditions at infinity. Within time-dependent
R-matrix theory, on the other hand, it takes time for the wavefunction to evolve, and
consequently, the boundary condition on the F-vector is that at a sufficiently
large distance the wavefunction, $F$, equals zero.

In order to obtain the time-dependent wavefunction, we need to consider the
outer-region wavefunction at a sufficiently large distance. Although Eq.
(\ref{eq:frtrelation}) is given at the inner-region boundary, it is a general
equation that holds throughout the outer region. The equation also
hold at a large distance where it can be assumed that the wavefunction, $F$, has
vanished. We thus
need to obtain the R-matrix and T-vector at this large distance. This is achieved
by dividing the outer region into subsectors, ranging from the inner-region
boundary $a_{r}$ out to this large distance $a_{p}$.
The Hamiltonian in each subsector is calculated in a similar way as in the
internal region by including Bloch operators $L_L$ and $L_R$ for the left-hand and right-hand
boundaries. We can then obtain the time-dependent Green's function for each subsector, and
use these Green's functions to propagate the R-matrix and T-vector from the inner-region
boundary $a_r$ to the outer boundary $a_p$ \cite{Burbook, Lys}.
Subsequently, we can use the R-matrix and
T-vector to propagate the F-vector inward from $a_p$ to $a_r$. Once we
have obtained $F$ across all subsector boundaries, we can determine the
wavefunction in the inner region and within all outer region subsectors, and initiate
the computation for the next time step. Repeating this procedure
at each time step, we can follow the behaviour of the wavefunction across the
full range of times. For more details on the propagation method, see \cite{Lys}. 

Calculation of the harmonic spectrum through the TDRM approach follows from determination of
dipole moment of the wavefunction at each time step.
The harmonic radiation emitted from the atoms and ions in an intense laser field
can be expressed in terms of the Fourier transform of the time-dependent expectation value
of either the dipole moment, the dipole acceleration
or the dipole velocity. The relative merits of each of these operators is still an active subject of discussion \cite{Bagg}.

 In the TDRM approach, we have a choice of using the length form,
\begin{equation}
  \mathbf {d}(t)=\langle\Psi(t)\vert -e\mathbf{z} \vert \Psi(t)\rangle,
\end{equation}  
or the velocity form,
\begin{equation}
  \mathbf {\dot {d}}(t)=\frac {d} {dt}\langle\Psi(t)\vert -e\mathbf{z} \vert\Psi(t)\rangle.    
\end{equation} 
The acceleration form is less appropriate for the TDRM approach, as restrictions on the basis
set mean that inner-shell electrons, such as the 1s electrons, are normally kept frozen. As
a consequence, the calculations include the action of the 1s electrons on valence electrons,
but the back-action on the 1s electrons is not taken into account. This limitation prevents
the use of the dipole acceleration in the determination of the harmonic spectrum. In the
present calculations, the harmonic spectrum is calculated using both the dipole length and
dipole velocity form, and we check for consistency between both spectra.

\subsection{Calculation Parameters}
 
As described above, basis functions for the description of Ne$^{+}$ states are expressed as
Ne$^{2+}$ residual-ion states plus an additional electron. We describe Ne$^{2+}$ using
Hartree-Fock orbitals for $1s, 2s$, and $2p$ of the Ne$^{2+}$ ground state, as given by
Clementi and Roetti \cite{Roe}. The R-matrix inner region has a radius of 15 a.u.
The continuum functions are described using a set of 60 B-splines of order $k=17$, 
for each available angular momentum $\ell$ of
the outgoing electron. We include all three $1s^{2}2s^{2}2p^{4}$ states, $ ^{3}P^{e}$,
$^{1}D^{e}$ and $^{1}$S$^{e}$, as residual-ion states. The description of Ne$^{+}$
includes all $1s^{2}2s^{2}2p^{4} n/\epsilon\ell$ Ne$^{+}$ channels up to a maximum total
angular momentum $L_{\rm max}$=23. In order to test the spectra for convergence some
calculations were also carried out for a angular momentum $L_{\rm max}$=27. 
The Ne$^+$ ground-state energy has not been shifted to its experimental value. 

In the TDRM calculations, the time step in
the wavefunction propagation for this calculation is normally set to 0.05 a.u. Additional
calculations were carried out at time steps of 0.04 a.u., and 0.06 a.u. with no significant
change in the overall spectrum. In the outer region we set the outer boundary to
1000 a.u. to prevent any unphysical reflections of the wavefunction.
The outer region is divided into subsectors of width 2 a.u. Here, the radial wave
function for each channel is described using a set of 35 B-splines of order 11. 
The laser pulse wavelength is chosen to be 390 nm. The pulse
profile is given by a three-cycle sin$^{2}$ turn-on followed by four cycles at peak intensity
and a three-cycle  sin$^{2}$ turn-off (3-4-3).

\section{Results}
In this report, we investigate HG from Ne$^+$ ions irradiated by laser light with
a wavelength of 390 nm. Ne$^{+}$ has been chosen for this current study due to its higher
ionization potential, 41 eV, compared to 27 eV for Ar$^{+}$. Due to the higher ionization potential,
the same level of ionization requires higher intensities, and it is thus possible to investigate
HG at higher intensities for Ne$^+$ compared to Ar$^+$. The higher intensity leads to an extended plateau
region for HG, and Ne$^+$ should therefore show in more detail how interference due
to channels associated with different ionization thresholds affects the harmonic spectra.

\begin{table}[!hbtp]
\caption {Energies of the three ionization thresholds of Ne$^{2+}$ with respect to the Ne$^{2+}$
ground state, and compared to literature values \cite{KR}.}
\begin{tabular*}{\columnwidth}{@{\extracolsep{\fill}}ccccc}
\hline
\hline
Configuration& Threshold &   Energy \cite{KR}  &    Energy (Present)    \\
				 &            & \footnotesize{eV}  & \footnotesize{eV}    \\ 
\hline
$2s^{2}2p^{4}$& $^ {3}P^{e}$& 0.00  & 0.00 \\ 
&$^{1}D^{e}$&3.20&3.42	   \\ 
&$^{1}S^e$&	6.91& 6.55 \\                                  
\hline
\end{tabular*}
\label{tab:thres}
\end{table}

The energies for the lowest three ionization thresholds of Ne$^{+}$, corresponding to the three different
$2s^{2}2p^{4}$ Ne$^{2+}$ states, as calculated in the present study, are listed in Table \ref{tab:thres}, and compared
to literature values. The three ionization thresholds are separated from each other by just over 3 eV. This
energy difference is comparable to the photon energy and, hence, the interplay between channels associated
with different thresholds cannot be neglected. The energy spacings in the present study differ
from the literature values, with the largest difference seen for the $^{1}D^e$ - $^{1}S^e$
gap which is 3.71 eV experimentally compared to 3.13 eV in the present study. The most important energy gap is the
$^{3}P^e$ - $^{1}D^e$ gap with a gap difference of 0.22 eV. These differences are sufficiently
small for identification of the most important effects of the interplay between channels.

%\begin{multicols}{
\begin{figure}[!hbtp]
\centering{\includegraphics[width=0.35\linewidth, angle=270]{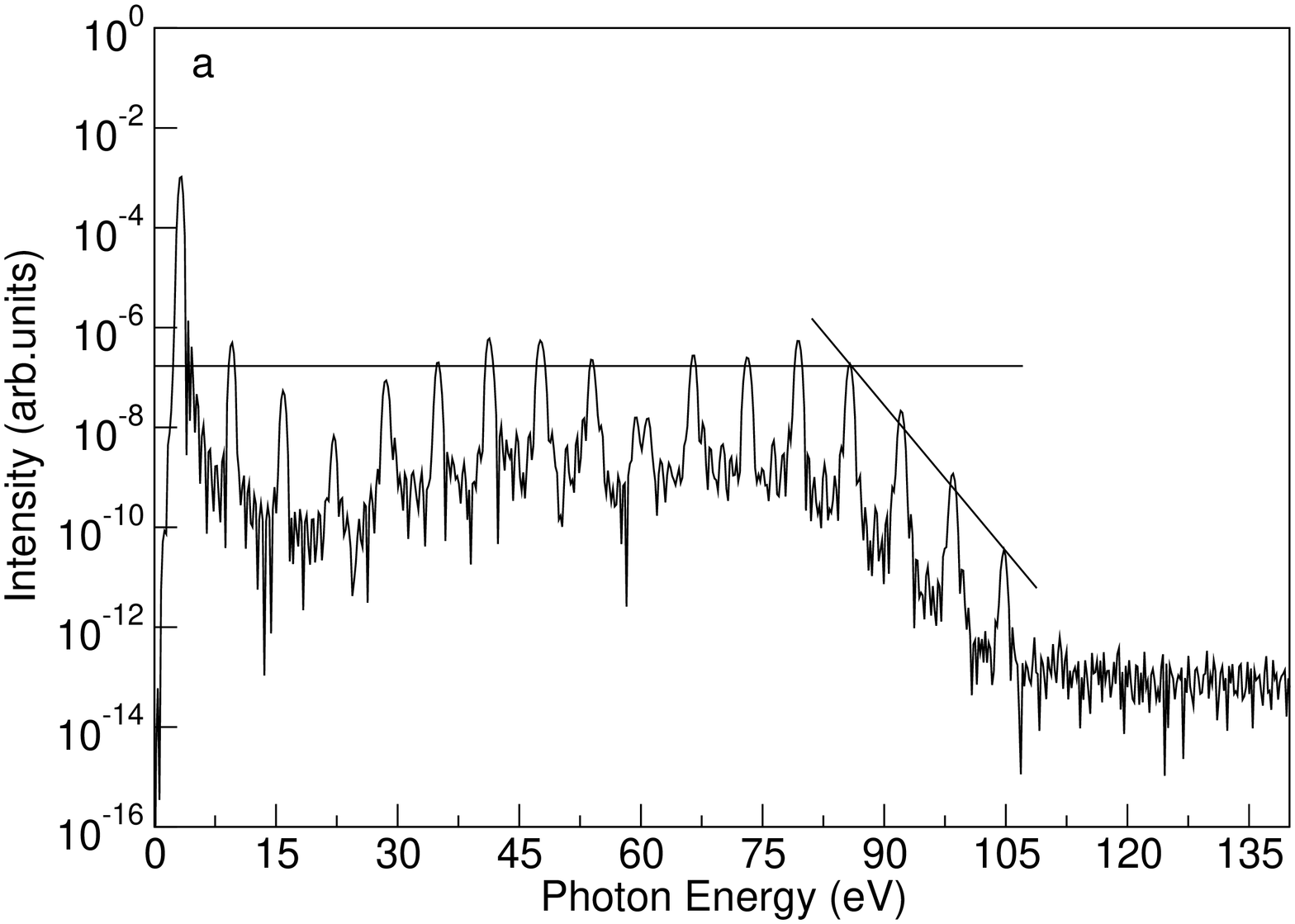}}\quad
\centering{\includegraphics[width=0.35\linewidth, angle=270]{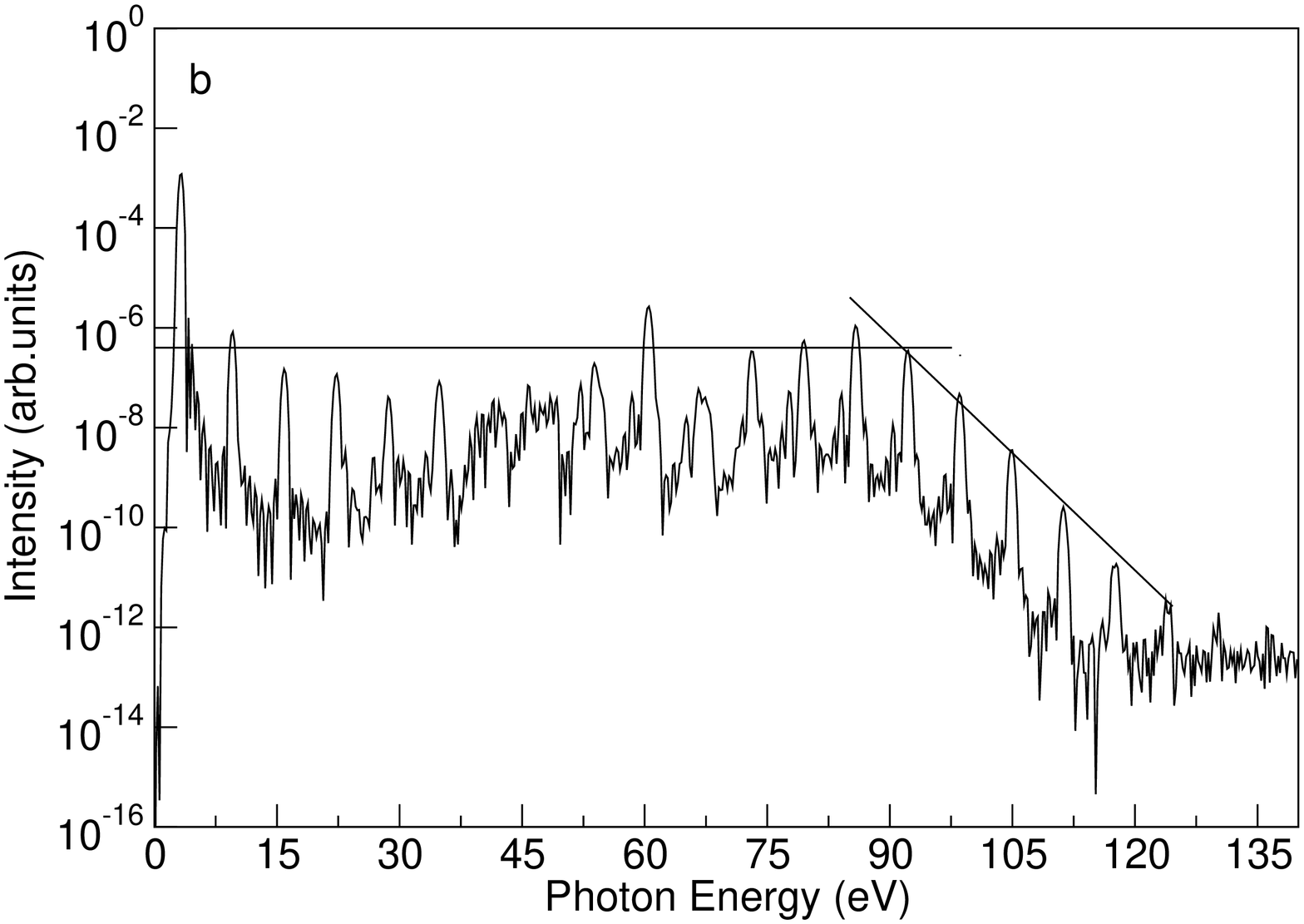}}\quad
\centering{\includegraphics[width=0.35\linewidth, angle=270]{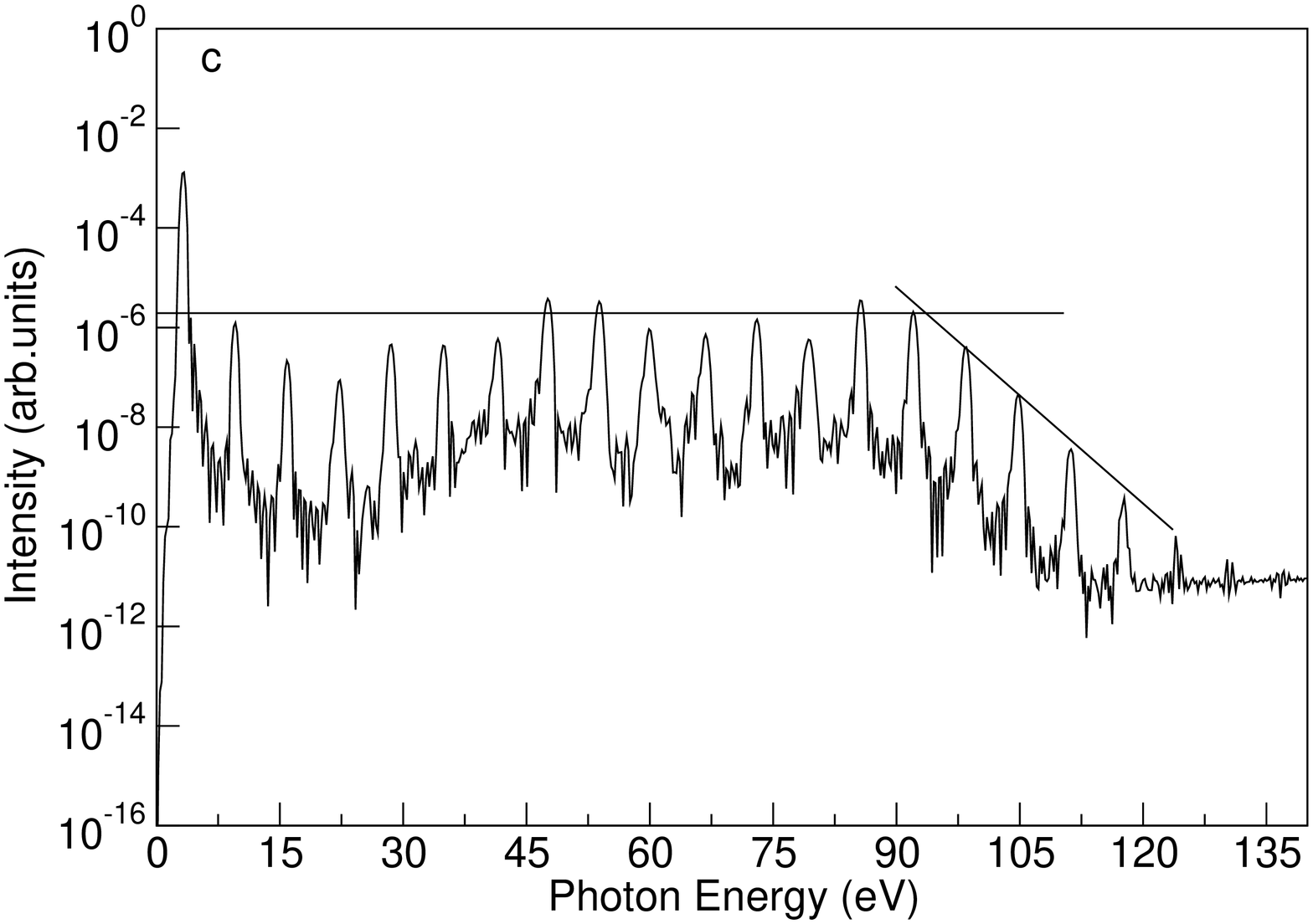}}\quad
\centering{\includegraphics[width=0.35\linewidth, angle=270]{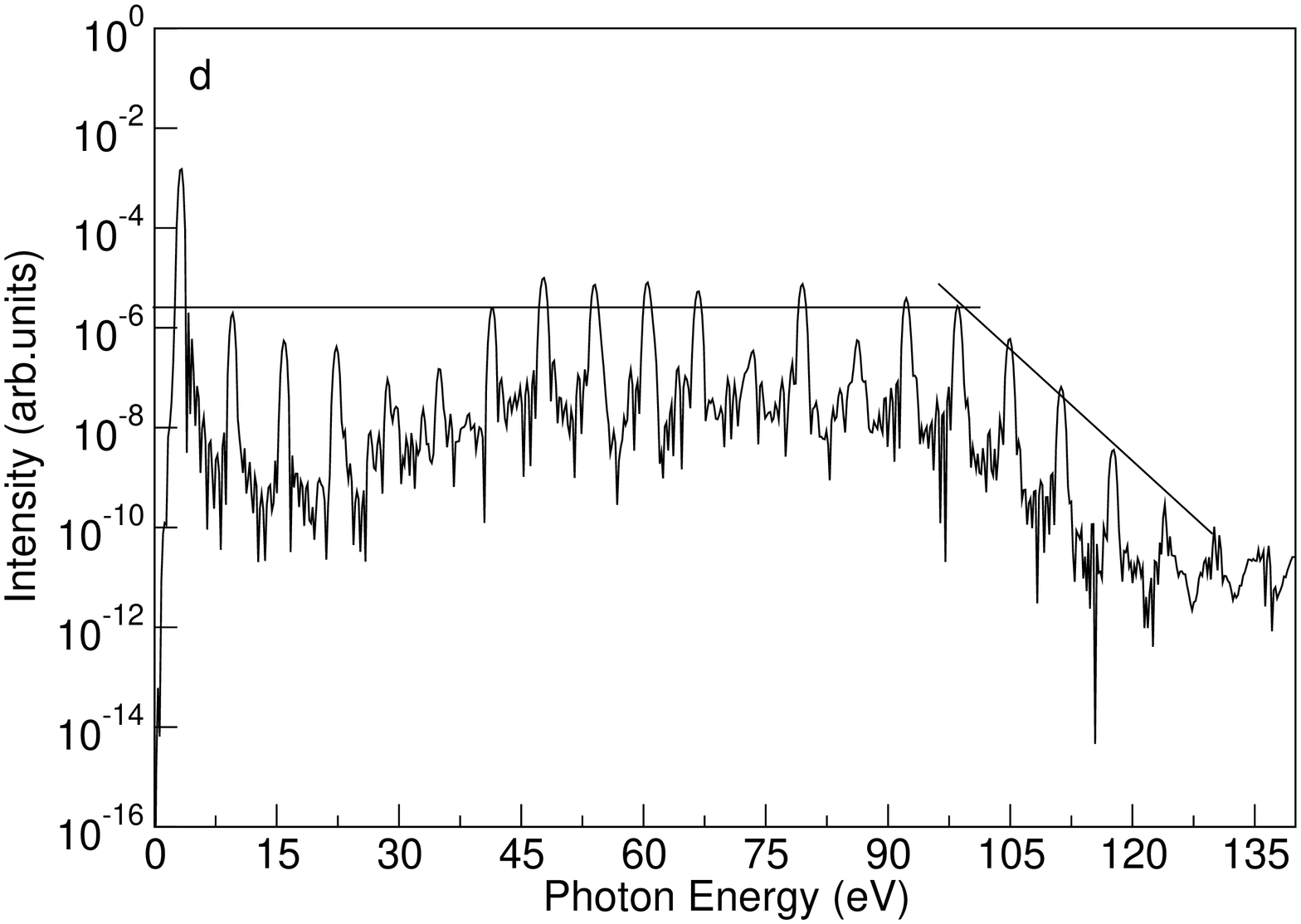}}
\caption{The harmonic spectrum obtained by 390 nm laser pulse with a total duration
of 10 cycles generated
from Ne$^+$ calculated from the dipole length, at different laser peak intensities:
(a) $7\times10^{14}$
Wcm$^{-2}$ (b) $8\times 10^{14}$ Wcm$^{-2}$ (c) $9\times 10^{14}$ Wcm$^{-2}$  and
(d) $10^{15}$ Wcm$^{-2}$.}
\label{fig:intens}
\end{figure}
%\end{multicols}

\subsection{High Harmonic generation from Ne$^{+}$ aligned with $M=0$}
\label{sec:m0}

The harmonic response of Ne$^{+}$, as calculated from the expectation value of the dipole length operator, is
shown in Fig. \ref{fig:intens}, for peak laser intensities between $7\times10^{14}$  Wcm$^{-2}$ and
$10^{15}$ Wcm$^{-2}$, and a 3-4-3 pulse profile. The spectra have the expected structure with a
plateau containing eight odd harmonics up to
a cut-off energy around 90 eV. Beyond this cut-off energy, an exponential decay in the harmonic yield is seen. 
The plateau is more extensive than observed for
Ar$^{+}$ at $4\times10^{14}$ Wcm$^{-2}$, for which the plateau contained 3 harmonic peaks \cite{Bro2,Bro3}.
Although the figure shows only the harmonic spectrum obtained through the dipole operator, the harmonic
spectrum has also been obtained using the expectation value of the dipole velocity.
The spectra calculated through the dipole and the dipole velocity show very good agreement. 
The convergence is typically within 20$\%$ in the overall harmonics up to 120 eV where the spectra associated from both
dipoles start to diverge clearly. The main reason for the overall difference is the limited
basis expansion used for the description of the Ne$^{2+}$ states in the present calculations.

Table \ref{tab:hyield} gives indicative intensities of the harmonic peaks, normalized to the harmonic
spectrum obtained at $10^{15}$ Wcm$^{-2}$, and the cutoff energy of the plateau as determined from the spectra.
The straight lines in Fig. \ref{fig:intens} demonstrate the origin of these values. 
Table \ref{tab:hyield} also gives the population in the outer region. 
In the three step model ionization is  the first step of HG.
The table shows that the increase in the harmonic yield approximately follows the increase in population in the outer region, with a factor 13 increase in the harmonic yield going from 7 $\times$10$^{14}$ Wcm$^{-2}$ to
10$^{15}$ Wcm$^{-2}$ matched by an increase of a factor 15 in the outer-region population.

\begin {table}[!hbtp]
\caption{Typical harmonic yield for Ne$^+$ irradiated by laser light with a wavelength of 390 nm,
normalized to the typical harmonic yield at a peak intensity of $10^{15}$ Wcm$^{-2}$, as a function
of peak intensity. The harmonic yield is calculated through the dipole operator, and the pulse
profile is a 3-4-3 pulse. The cut-off energy of the harmonic plateau is given as well and compared with the
prediction of the cut-off formula, $1.25I_p + 3.17 U_p$. The final population in the outer region is
given as well.}
\begin{tabular*}{\columnwidth}{@{\extracolsep{\fill}}ccccc}
  \hline
  \hline
Intensity  &     Relative harm-   & Outer region &   Cut-off          &   Cut-off       \\
           &       onic yield    &          population  &       &   \footnotesize{$1.25I_p + 3.17U_p$} \cite{Le}   \\
%------------------------------------------------------------------------------			  
\footnotesize{ (Wcm$^{-2}$)} & \footnotesize{(eV)}  & \footnotesize{($10^{-5}$)}&  \footnotesize{(eV)}&    \footnotesize{(eV)}\\ 
\hline
$1.0\times 10^{15}$  & 1.0\phantom{0}  &  8.43 &98.2& 96.4 \\ 
$0.9\times 10^{15}$  & 0.29            &  4.14 &93.8 & 91.9\\ 
$0.8\times 10^{15}$  & 0.17            &  1.72 &92.5 &  87.4\\
$0.7\times 10^{15}$  & 0.08            &  0.58 & 86.2 & 82.8\\  
\hline
\end{tabular*}
\label{tab:hyield}
\end{table}

Table \ref{tab:hyield} shows the variation of the cut-off energy of the plateau with peak intensity.
The determination of the cut-off energy is shown in the graphs. These cut-off values are expected to
have an energy uncertainty $\leq$ 1.5 eV.
The standard cut-off formula for the energy of the cut-off is given by $\alpha I_{p} +U_{p}$ \cite{Le},
where $U_{p}$ is the ponderomotive potential of a free electron in a laser field, $I_{p}$ is
the ionization potential and $\alpha$ is a parameter, which depends on the ratio between $I_p$ and $U_p$. For the present range of intensities, the parameter ranges between 1.235 and 1.25. For simplicity, we adopt $\alpha=1.25$. Values obtained from this formula are also given in the table. It can be seen that for all intensities, the observed cut-off energy values differ by about 2 eV from the predictions of the formula, although at an intensity of $0.8\times 10^{15}$ Wcm$^{-2}$ the cut-off formula underestimates the observed cut-off energy by 5.1 eV.

Figure \ref{fig:intens} shows great variation of peak intensity within the
plateau region. In Fig.
\ref{fig:intens}(b), harmonic 19 at 60 eV is a factor 40 more intense than harmonic 21. In Fig. 
\ref{fig:intens}(d), relatively little harmonic response is observed for harmonics 23 and 27, at 72 eV
and 86 eV, respectively. The basis set in the present calculations is chosen specifically to exclude
resonances above the $2s^{2}2p^{4}$  $^{1}S^{e}$ ionization threshold, so that this reduction in
magnitude cannot ascribed to atomic structure at these harmonic energies. Harmonics up to a photon
energy of around 50- 55 eV, on the other hand, can be strongly affected by resonances due to the
Rydberg series leading up to the $2s^{2}2p^{4}$ thresholds.

The HG spectra can be affected by interferences arising from several sources. Below the cut-off energy, harmonics can be created by electrons returning on so-called short and long trajectories \cite{Sm}. For multi-threshold systems, interferences between channels associated with different thresholds can arise as well. Within the TDRM approach, it is difficult to unambiguously separate short and long trajectories. Therefore, in order to understand whether the
interplay between channels associated with different thresholds could be responsible for the interference
structure, as well as the discrepancy in the cut-off energy, we have carried out additional
calculations in which only subsets of the $2s^{2}2p^{4}$ thresholds are taken into account.

%\begin{multicols}{
\begin{figure}[!hbtp]
\centering
\includegraphics[width=0.35\linewidth, angle=270]{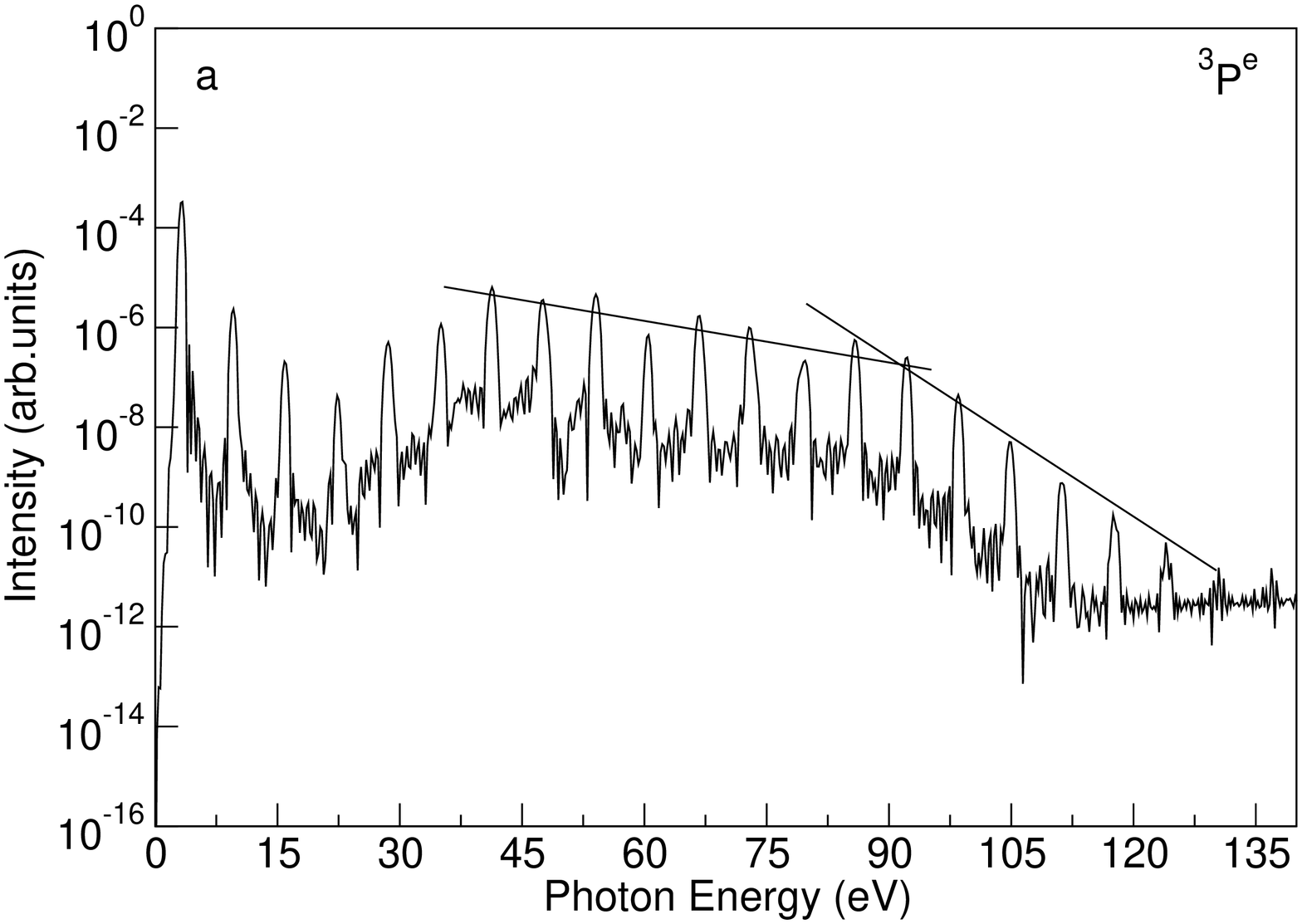}
\includegraphics[width=0.35\linewidth, angle=270]{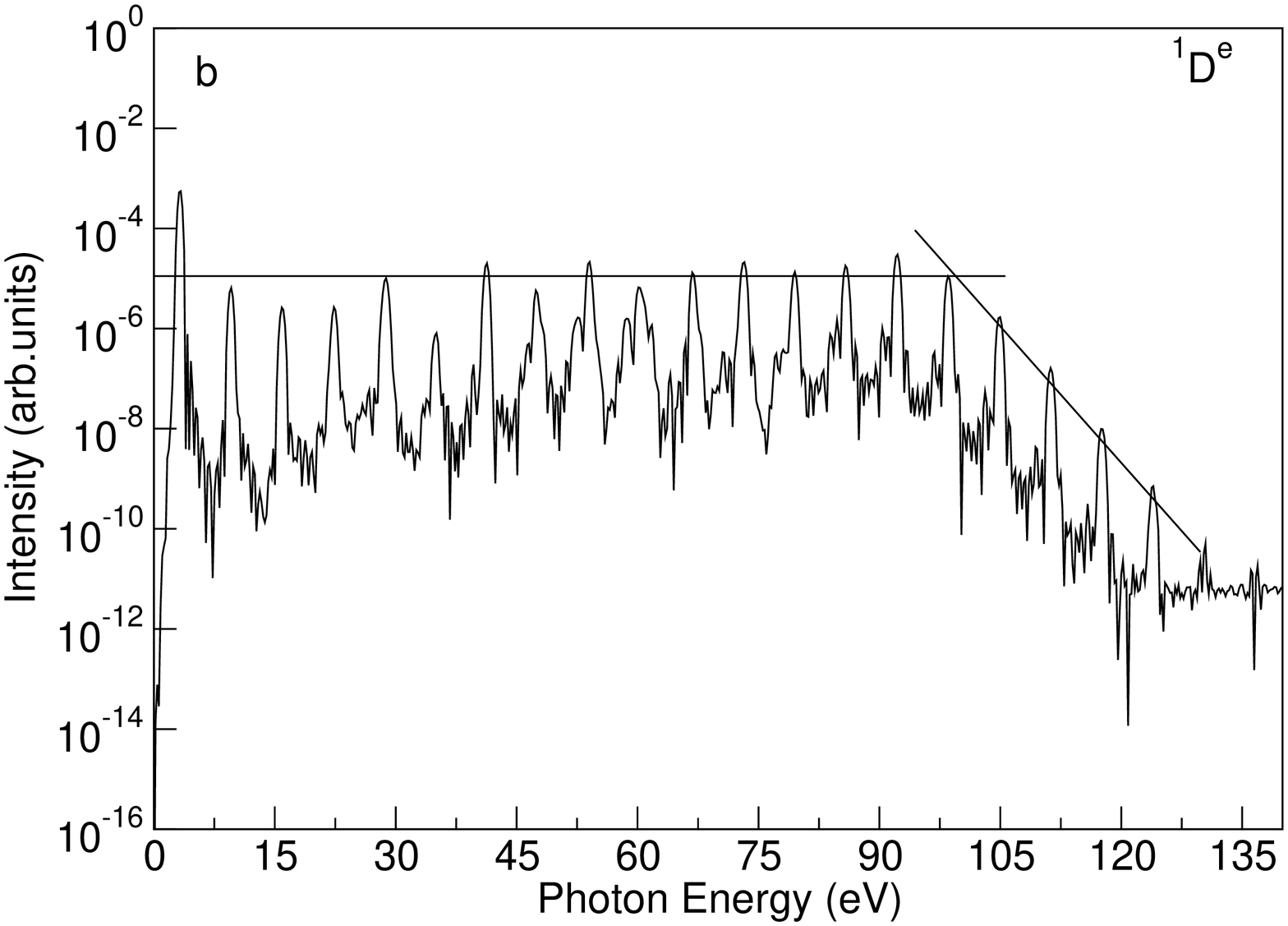}
\includegraphics[width=0.35\linewidth, angle=270]{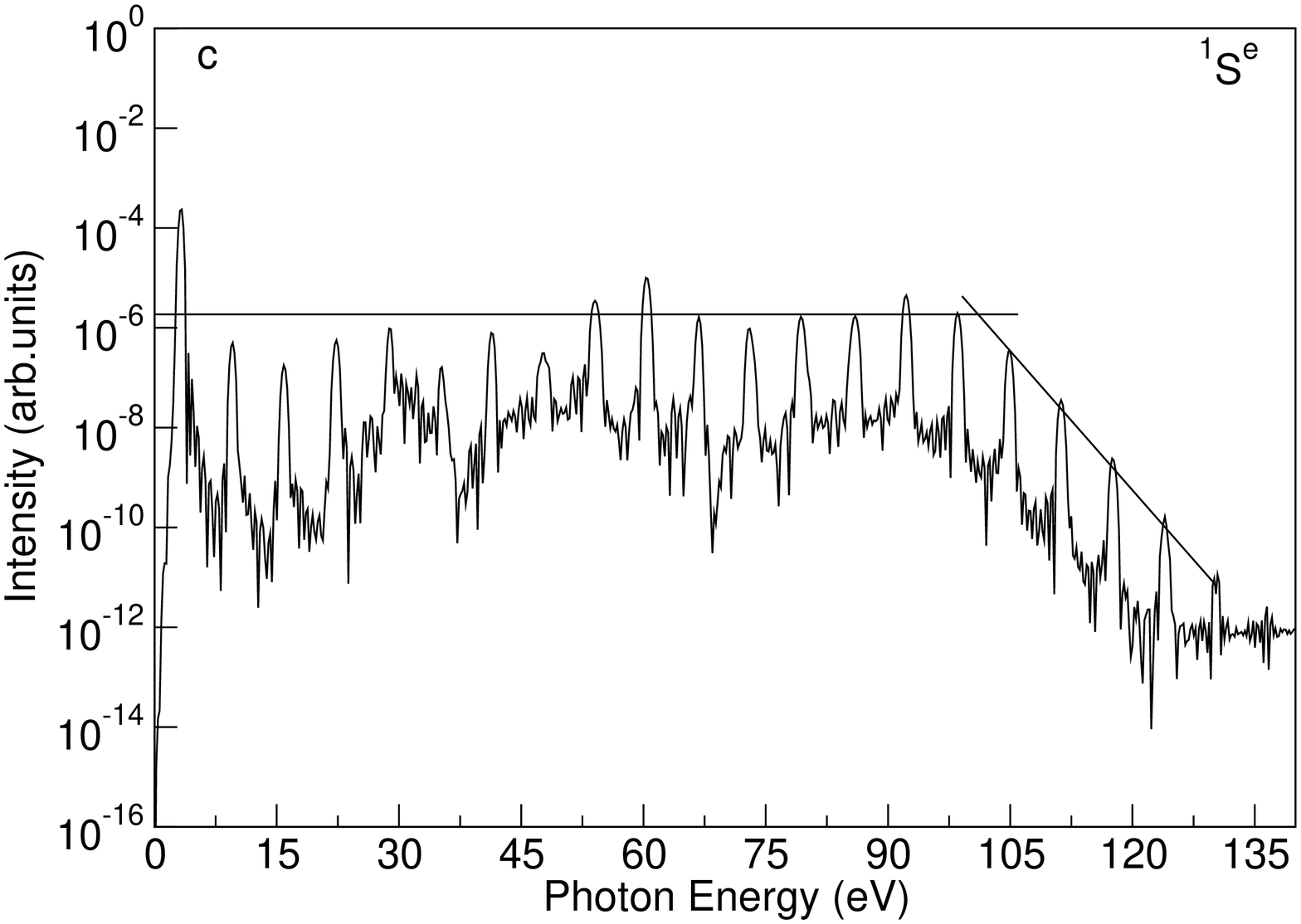}
\includegraphics[width=0.35\linewidth, angle=270]{allm0.ps}
\caption{Harmonic spectra for Ne$^+$ irradiated by 390 nm laser pulse as calculated through
the dipole operator, at peak intensity of $10^{15}$ W cm$^{-2}$. The Ne$^{2+}$ 2s$^2$2p$^4$ 
residual-ion states retained are: (a) $^{3}P^e$,(b) $^{1}D^e$, (c) $^{1}S^e$,  and
(d) all three states.}
\label{fig:single}
\end{figure}
%}
%\end{multicols}

First, we consider the harmonic spectra when only a single residual ion state is included
in the calculation. Since harmonics 23 and 27 show a significant reduction in magnitude
for an intensity of $10^{15}$ W cm$^{-2}$,  we have chosen this intensity for the comparison.
Figure \ref{fig:single} shows the harmonic spectra
when only an individual $2s^{2}2p^{4}$ threshold of Ne$^{2+}$, ($^{3}P^e, ^{1}D^e$ or $^{1}S^e$) is
included. For comparison, the figure also includes the full spectrum.

Figure \ref{fig:single} shows significant variation in the harmonic efficiency between
the different individual threshold calculations. In the calculation in which only the
$^{1}D^{e}$ threshold is retained,
the observed harmonic intensities are about an order of magnitude greater than those observed
when only the $^{1}S^e$ threshold is retained. The harmonic intensities when only the
$^{3}P^{e}$ threshold, the Ne$^{2+}$ ground state, is retained are about one order of magnitude
smaller than those when only the $^{1}D^{e}$ threshold is retained
at 45 eV, and about two orders of magnitude smaller at 90 eV. The spectrum obtained
when all thresholds are included shows harmonic intensities one order of magnitude smaller than the
spectrum obtained when only the $^{1}D^{e}$ threshold is retained. This is a significant
variation in harmonic yields depending on the symmetry of the ionization threshold. Since the
smallest harmonic yields are retained when only the Ne$^{2+}$ ground state is obtained, this
variation cannot be explained solely by the variation in binding energy of the
different thresholds.

\begin {table}[!hbtp]
\caption{Typical harmonic yields for Ne$^+$ irradiated by 390 nm laser light at an intensity of 10$^{15}$
Wcm$^2$ as a function of thresholds retained, normalized to the spectrum obtained when all three thresholds
are included. The cut-off energy of the plateau region and the final outer region population are also shown
for each subset of Ne$^{2+}$ thresholds retained. No typical harmonic yield can be given when only
the $^3P^e$ threshold is retained (see text).}
\begin{tabular*}{\columnwidth}{@{\extracolsep{\fill}}lccc}
\hline
\hline
Threshold  & Relative       &     Cut-off           &     Population \\
           & harmonic yield & \footnotesize{(eV)}   &  in outer region  \\ 
\hline
$^{1}D^{e}$,$^{3}P^{e}$ and $^{1}S^{e}$ & 1\phantom{.00}  & 98.1 & $8.38\times10^{-5}$ \\                         
$^{1}D^{e}$                             & 4.1\phantom{0}  &	98.6 & $3.23\times10^{-4}$ \\ 
$^{3}P^{e}$                             &      				 &	92.4 & $8.89\times10^{-5}$ \\
$^{1}S^{e}$                             & 0.64 				 &	99.1 & $6.02\times10^{-5}$ \\  
$^{1}D^{e}$,$^{3}P^{e}$                 & 2.1\phantom{0}  &	99.2 & $9.07\times10^{-5}$ \\ 
$^{3}P^{e}$,$^{1}S^{e}$                 & 0.28 				 &	97.2 & $7.52\times10^{-5}$ \\ 
$^{1}D^{e}$,$^{1}S^{e}$                 & 4.4\phantom{0}  & 98.2 & $2.77\times10^{-4}$ \\
\hline
\end{tabular*}
\label{tab:indiv}
\end{table}

As demonstrated earlier, the typical harmonic yield behaves similarly to the population in the
outer region. We focus our attention therefore first on these populations, shown in
Table \ref{tab:indiv}. Note that the table does not provide a typical harmonic yield when only the $^{3}P^e$
threshold is retained. Figure \ref{fig:single}(a) shows a decrease of about an order of magnitude
across the plateau region, and it is therefore impossible to identify a typical value that applies
to the entire plateau. Table \ref{tab:indiv} shows that the variation in the typical harmonic yields
is reflected to some extent in the outer-region populations. The outer region population when only
the $^1D^e$ threshold is retained is about a factor 3.7-5.5 larger than when either the $^{3}P^e$ or
$^1S^e$ threshold is retained. It is, however, also a factor 4 larger than when all three
thresholds are retained. The increase in the population in the outer region may therefore
be the fundamental reason for the higher harmonic yields when only the $^{1}D^e$ threshold is retained. 
However, the population in the outer region does not explain the difference between the
harmonic yields obtained when only the $^{1}S^e$ or the $^{3}P^e$ threshold is retained.

Table \ref{tab:indiv} also shows how the cut-off energies for the plateau region depend
on the thresholds retained in the calculation. The cut-off energies when only the
$^{1}D^e$ or $^{1}S^e$ threshold is retained lie within 1 eV of the cut-off energy
observed when all three thresholds are retained. However, the cut-off energy obtained
when only the $^{3}P^e$ threshold is retained is nearly 6 eV smaller than the cut-off
energies obtained in the other calculations. This is clear demonstration that there are
fundamental differences between the process of HG in Ne$^+$ with $M=0$
when all three thresholds are accounted for and when only the Ne$^{2+}$ ground state
is accounted for.

Figure \ref{fig:single}(d) shows the full harmonic spectrum, in which harmonics 23 and 27
are reduced in intensity by about 1.5 order of magnitude. However, Fig. \ref{fig:single}(a), (b)
and (c), show no significant reduction in intensity at either of these harmonics.
Instead, the spectra show smooth variations in the intensities of the
different harmonic peaks. These single-threshold spectra are still affected by interferences between the short and long trajectories. No sign of these interferences is seen in the individual spectra, although there is some variation in the magnitude of individual harmonic peaks. Hence, the reduction in the harmonic intensity seen in Fig. \ref{fig:single}(d)
cannot be assigned to dynamics associated with a single ionization threshold. It is therefore
necessary to consider harmonic spectra obtained when multiple ionization thresholds are retained.

%\begin{multicols}{
\begin{figure}[!hbtp]
\centering
\includegraphics[width=0.35\linewidth, angle=270]{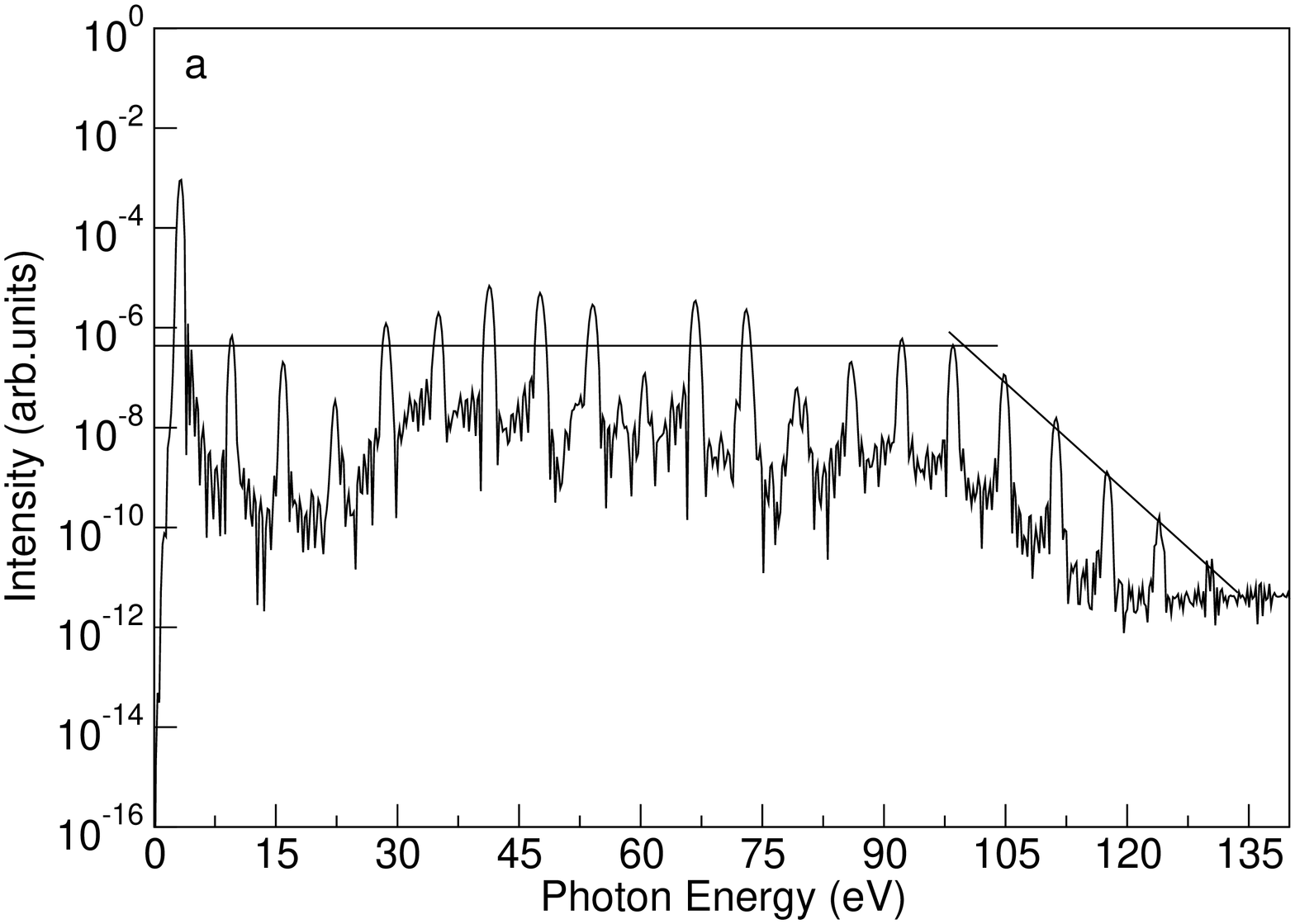}
\includegraphics[width=0.35\linewidth, angle=270]{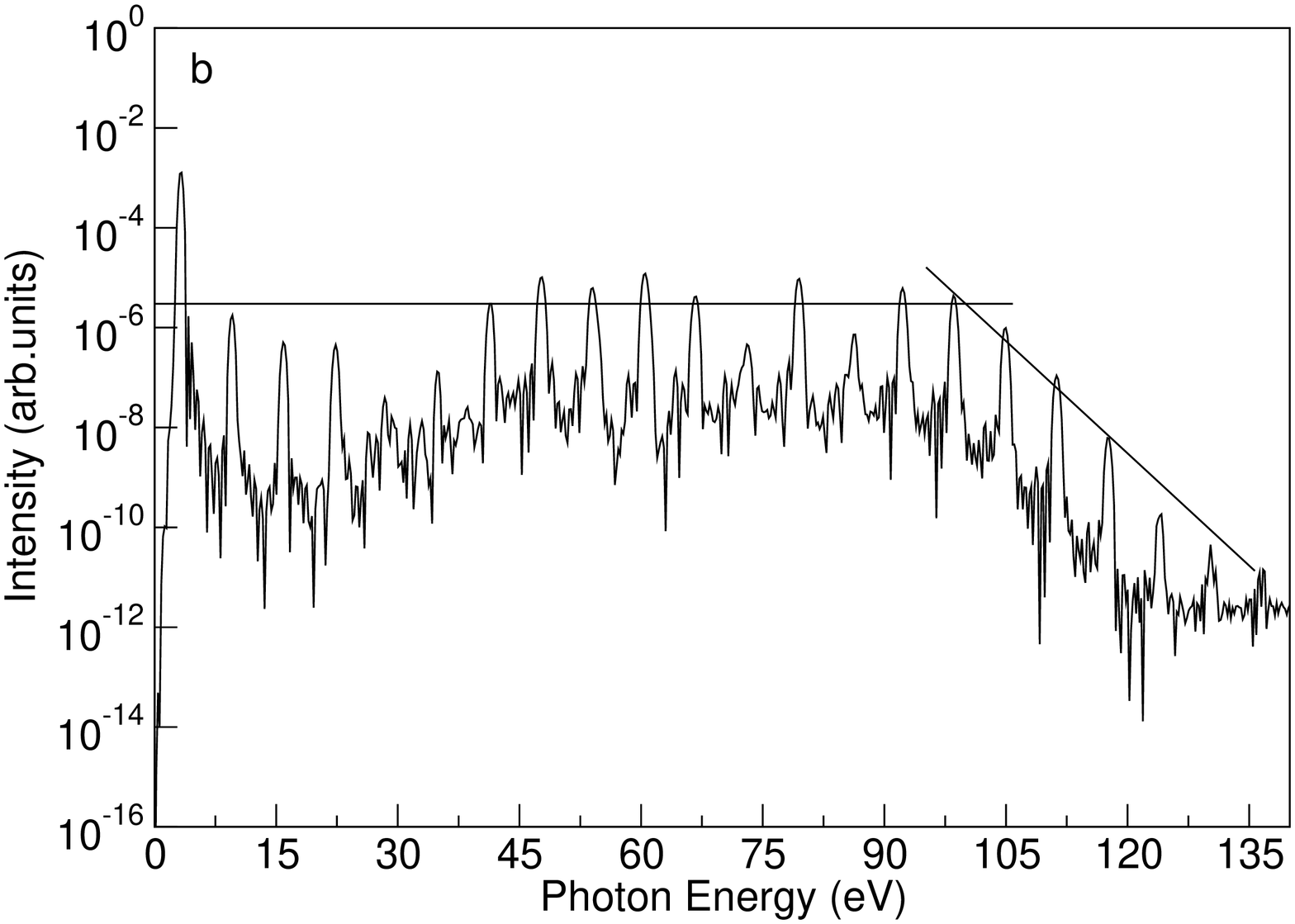}
\includegraphics[width=0.35\linewidth, angle=270]{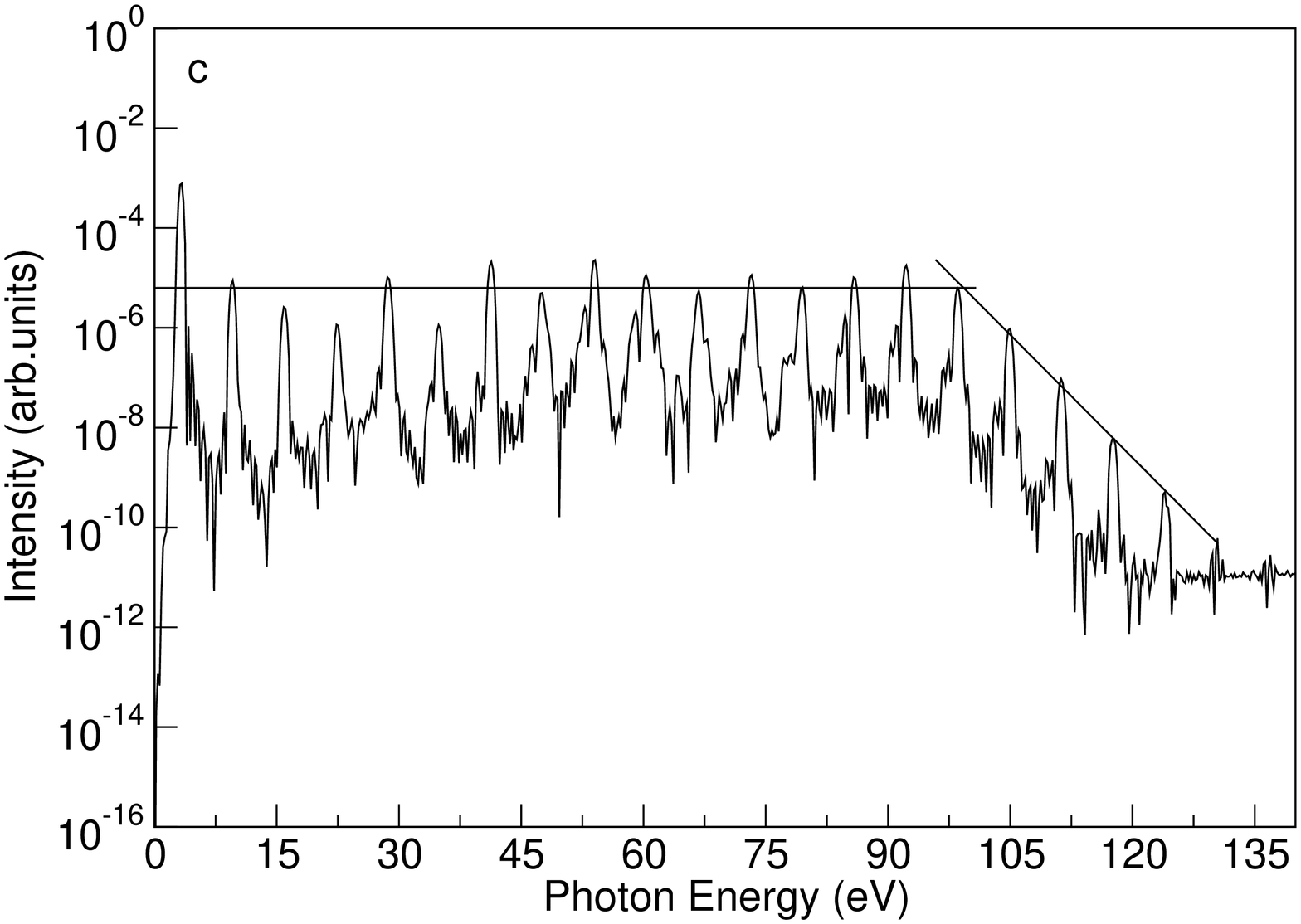}
\includegraphics[width=0.35\linewidth, angle=270]{allm0.ps}
\caption{The harmonic spectra obtained for Ne$^+$ irradiated by 390 nm laser pulse,
as obtained through the dipole length operator, at peak intensity of $10^{15}$ W cm$^{-2}$.
The spectra are obtained when the following 2s$^2$2p$^4$ thresholds are retained: 
(a) $^{3}P^e$ and $^{1}S^e$, (b) $^{3}P^e$ and $^{1}D^e$, (c) $^{1} S^e$ and
$^{1} D^e$ thresholds and (d) all three thresholds.}
\label{fig:pairs}
\end{figure} 
%}
%\end{multicols}

The next stage in the analysis is to investigate harmonic spectra when pairs of Ne$^{2+}$
residual-ion states, ($^{1}D^e$, $^{3}P^e$), ($^{1}S^e$, $^{3}P^e$) and ($^{1}D^e$,
$^{1}S^e$), are
retained in the calculations. These harmonic spectra are shown in Fig. \ref{fig:pairs}
and compared to the spectrum obtained in the full calculation. The harmonic spectrum obtained
when both the $^{1}D^e$ and $^{3}P^e$ thresholds are retained shows
great similarity to the harmonic spectrum obtained when all three target states are retained.
The other spectra show noticeable differences with the full spectrum. 
The spectrum obtained when both the $^{3}P^e$ and  $^{1}S^e$ thresholds are retained
shows harmonic intensities, which are about a factor of 4 smaller than those of the full spectrum,
especially at higher energies. The ($^{1}D^e$, $^{1}S^e$) spectrum has harmonic intensities
which are slightly larger than the full spectrum in the plateau region. However, the yield
for harmonics below the ionization threshold is larger than obtained in the full calculation
and these peaks are more pronounced compared to the full spectrum. No sign of interference
is seen for harmonics 23 and 27. 

Once again, Table \ref{tab:indiv} lists the population in the outer region when each pair of
Ne$^{2+}$ residual-ion states ($^{1}D^e$, $^{3}P^e$), ($^{1}S^e$, $^{3}P^e$) and
($^{1}D^e$, $^{1}S^e$) is
retained in the calculation. When the $^{3}P^e$ threshold is included in the calculation, the total
population in the outer region is within 10\% of the outer-region population when all three thresholds are
included. However, when only the $^{1}D^e$ and $^{1}S^e$ thresholds are accounted for, the population
in the outer region is larger than the population in the full calculation by more than a factor 3. Hence
the inclusion of the $^{3}P^e$ threshold is essential to obtain an accurate population in the outer region.

The cut-off energies, given in Table \ref{tab:indiv}, show agreement with the full spectrum within
1 eV. The interaction between channels associated with the $^3P^e$ threshold and channels associated
with either the $^{1}D^e$ or $^{1}S^e$ threshold leads to the cut-off energy of the harmonic spectrum being
shifted upward by about 6 eV.

Signs of destructive interference can be observed in the spectra obtained in both the ($^{3}P^e$, 
$^{1}S^e$) and ($^{3}P^e$, $^{1}D^e$) calculation. In the former case,  harmonics 25 and 27 are suppressed,
 whereas harmonics 23 and 27 are suppressed in the latter case. For the ($^{3}P^e$, $^{1}D^e$) case, the decrease 
in magnitude corresponds well to the decrease observed in the full calculation.

Overall, these harmonic spectra demonstrate that HG from ground-state Ne$^+$ with $M=0$ at a
wavelength of 390 nm requires the inclusion of at least the 2s$^2$2p$^4$ $^{3}P^e$ and $^{1}D^e$
thresholds of Ne$^{2+}$.  The harmonic spectrum is dominated by the harmonic response of the
excited $^{1}D^e$ threshold, as demonstrated, for example, by the cut-off energy in the full calculation.
However, if just the $^{1}D^e$ threshold is accounted for, the ionization rate is too high, and the
spectra do not show the right level of variation in harmonic intensities. Inclusion of
the (lower-lying) $^{3}P^e$ threshold reduces the ionization rate, and the
correct magnitude of the harmonic
yield is obtained. Interference between pathways associated with these ionization thresholds then
leads to the destructive interference for certain harmonics.

It is counterintuitive that inclusion of a lower-lying ionization threshold reduces the ionization
rate. The lower ionization rate associated with the $^{3}P^e$ threshold can be explained through
the allowed $m$-values of the ejected electron \cite{Bro3}. For $M=0$,
ejection of an electron  with $m=0$ is only allowed for the $^{1}D^e$ threshold. The $^{3}P^e$ threshold can
only be reached through the emission of $m=1$ electrons. The Rydberg series converging to the
$^{1}D^e$
and the $^{3}P^e$ thresholds overlap. When an $m=0$ electron is excited towards the $^{1}D^e$
threshold,
electron-electron interactions between the two channels can `re-route' an electron from the
$^{1}D^e$ path
to the $^{3}P^e$ path changing the $m$-value of the electron. Since below the $^3P^e$ state
the density of states in
the Rydberg series converging to the $^{3}P^e$ threshold is higher than that for the series converging
to the $^{1}D^e$ threshold, it may be difficult for the electron to return from the $^3P^e$ Rydberg series
to the $^{1}D^e$ Rydberg series. This process can slow
down ionization and, through reduction of the $^{1}D^e$ ionization pathway, reduce the 
harmonic yield. 

\subsection{High Harmonic generation from Ne$^{+}$ aligned with $M=1$}
\label{sec:m1}

%\begin{multicols}{
\begin{figure}[!hbtp]
\centering
\includegraphics[width=0.35\linewidth, angle=270]{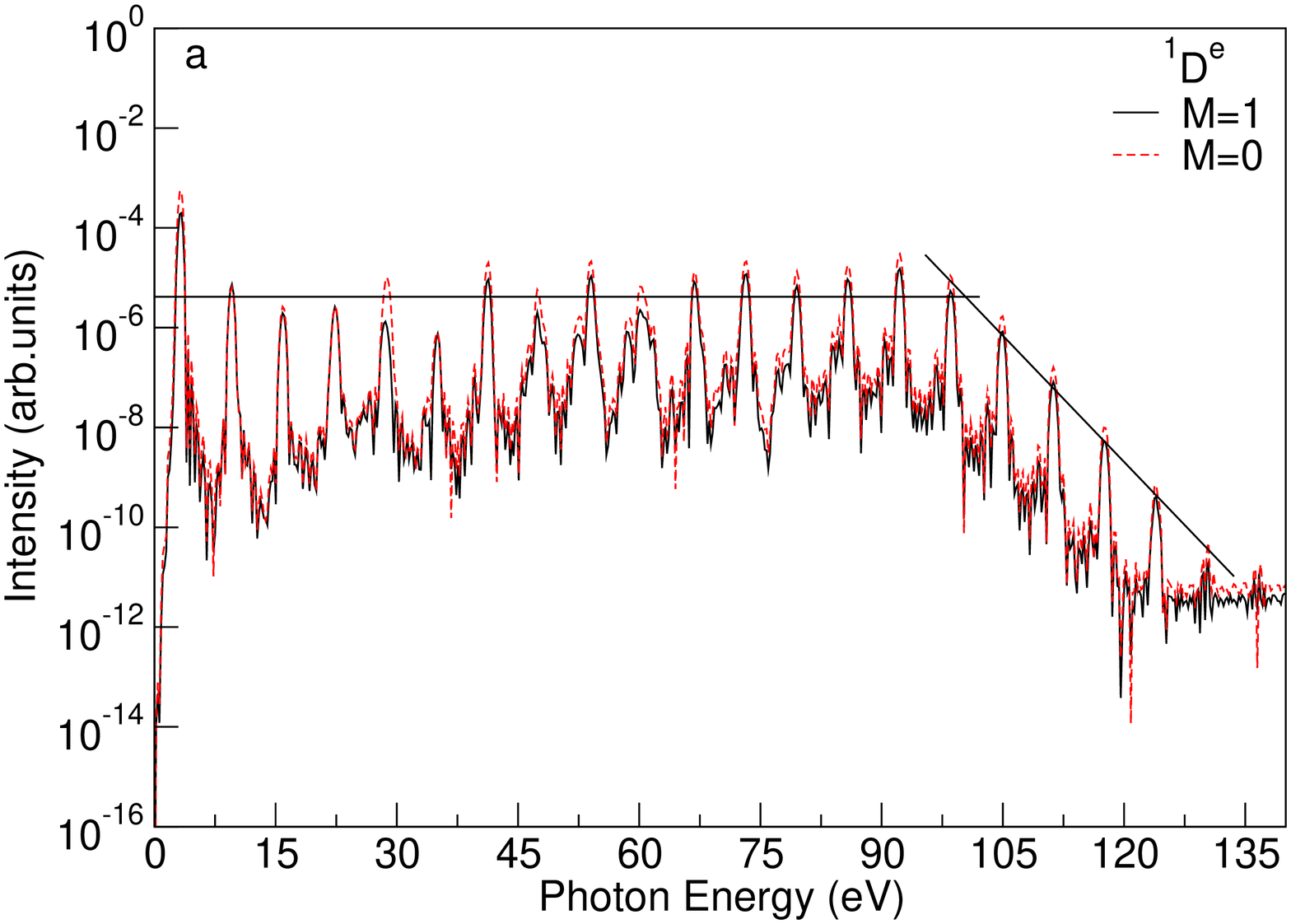}
\includegraphics[width=0.35\linewidth, angle=270]{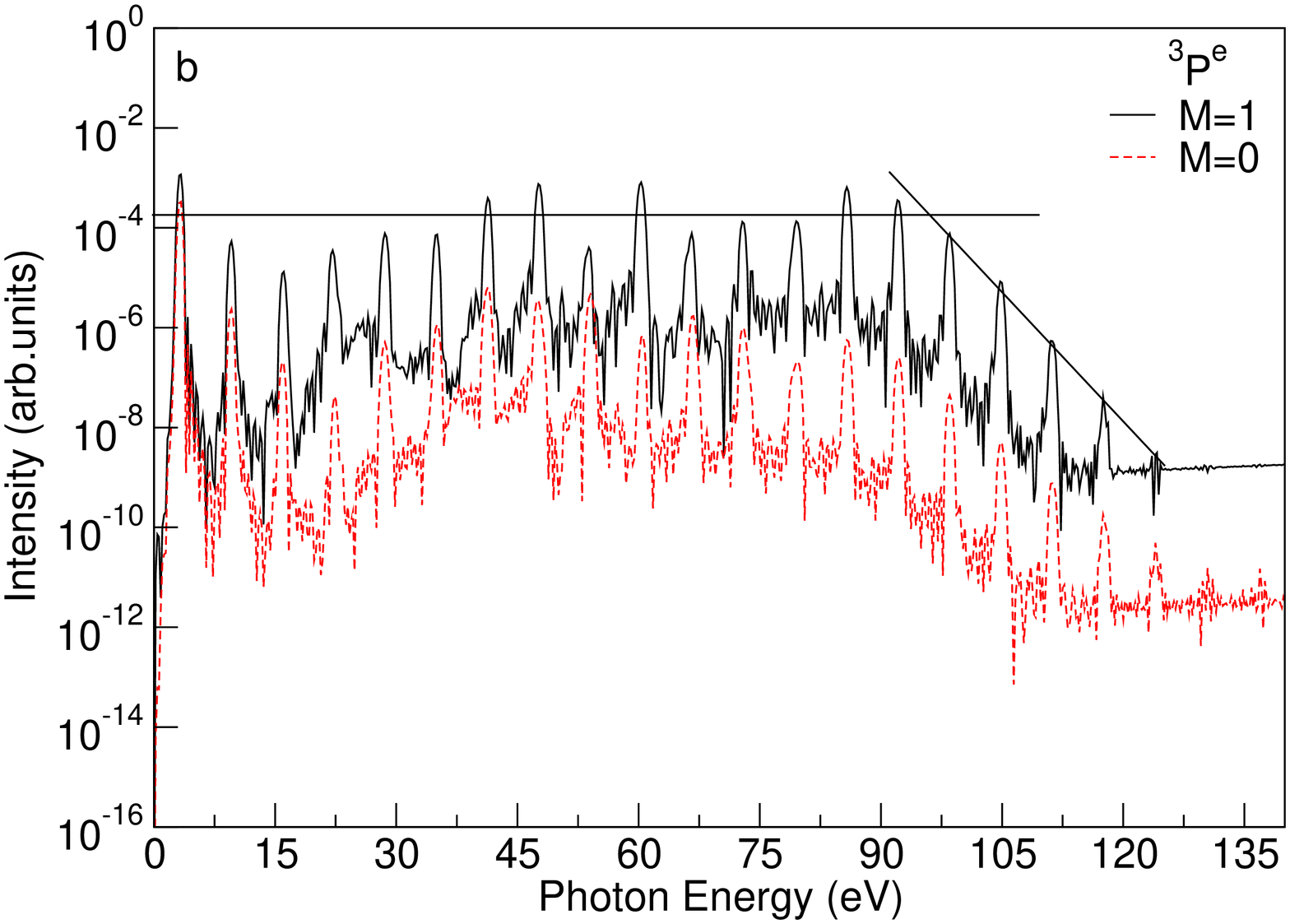}
\includegraphics[width=0.35\linewidth, angle=270]{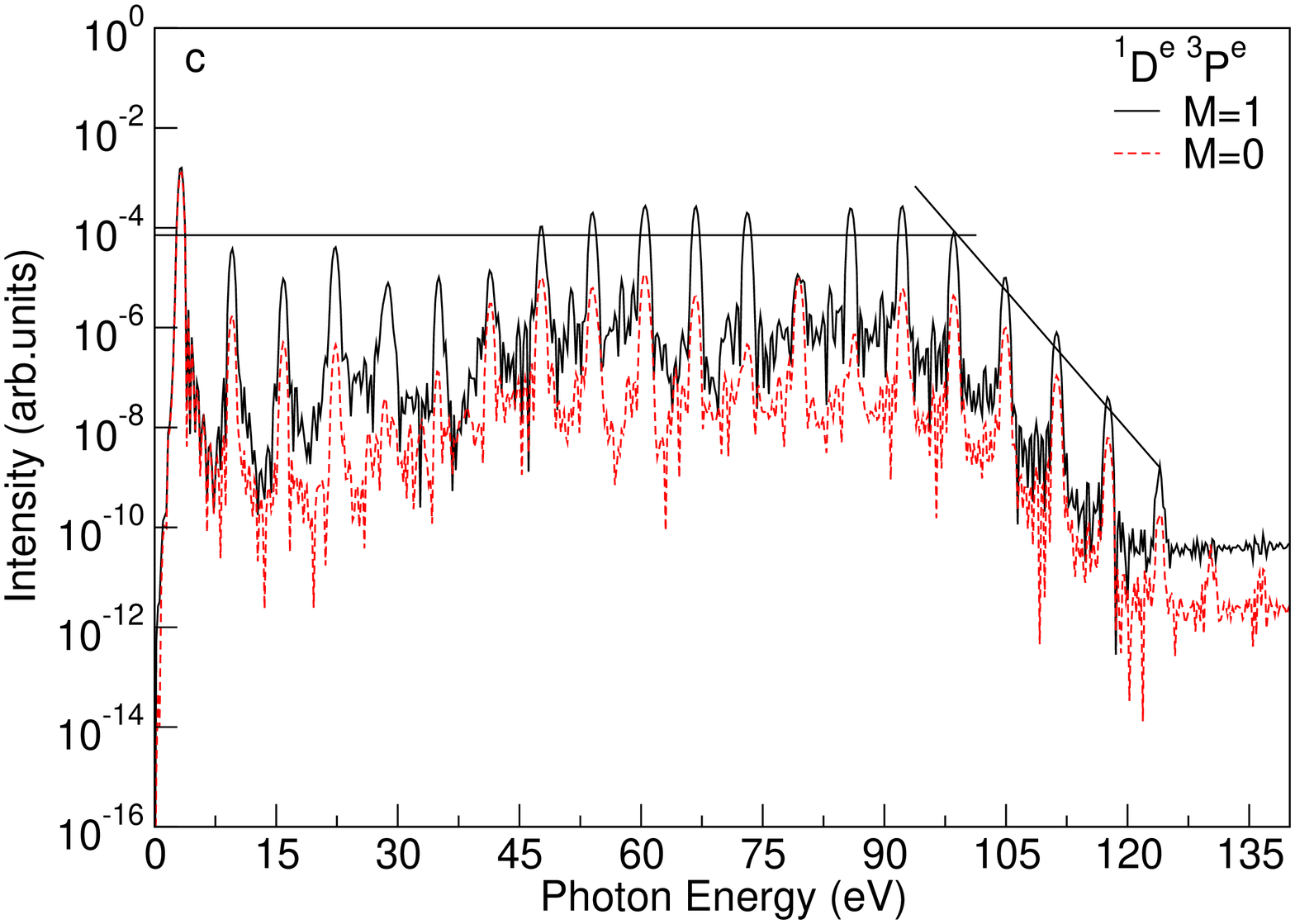}
\includegraphics[width=0.35\linewidth,angle=270]{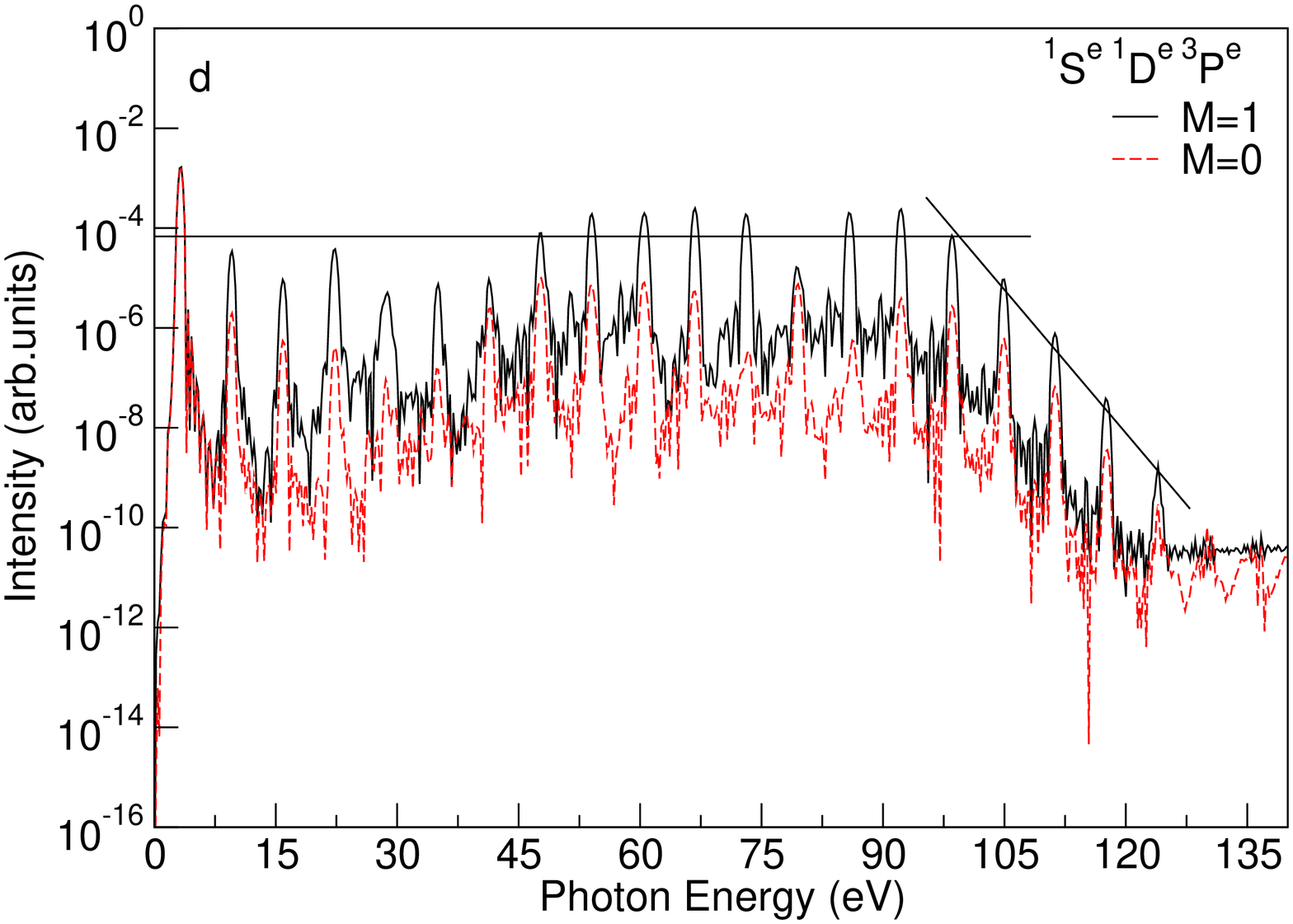}
\caption{(Color online) The harmonic spectra obtained by 390 nm laser pulse
with a total duration of 10 cycles generated from Ne$^+$ as calculated from
the dipole length, at peak intensity of $10^{15}$ Wcm$^{-2}$ with different
thresholds for both $M=1$ (solid-black line) and $M=0$ (dashed-red line): (a) $^{1}D^e$,(b) $^{3}P^e$,
(c) $^{1}D^e$ and $^{3}P^e$ and (d) the full configuration.}
\label{fig:m1}
\end{figure}
%}
%\end{multicols}

In the previous study of HG in Ar$^+$, a difference of about a factor of 4 was
observed between the harmonic yields for $M=0$ and $M=1$. It is therefore valuable to investigate
whether a similar difference is observed for Ne$^+$ as well, and to see whether the magnetic quantum
number leads to changes in the interference pattern.
The total magnetic quantum number $M$ has a significant effect on the calculations: it affects
the allowed radiative transitions. For systems with $M=0$, the selection rules state that only
radiative transitions with $\Delta L =\pm1$ are allowed, but, for $M=1$,
$\Delta{L} = 0, \pm1$ radiative transitions are allowed. Hence, transitions between states
with different parity but the same angular momentum ($>0$) are now allowed. This doubles 
the number of total symmetries that need to be retained in the calculation, with
a corresponding increase in the size of the calculations. We have therefore performed only a
limited number of calculations for $M=1$, focussing primarily on the interplay between the
$^3P^e$ and $^1D^e$ thresholds.

Figure \ref{fig:m1}(d) shows the harmonic spectrum for Ne$^+$, with $M=1$ initially, obtained when all
three residual Ne$^{2+}$ states are retained in the calculation. The $M=0$ spectrum is included
for comparison. The figure shows an increase for $M = 1$ of about a factor 26 compared to $M = 0$.
This reflects, in part, an increase in the population in the outer region by a factor of 12.5 for $M = 1$ 
compared to $M = 0$. This change in the harmonic yield is a factor 6 larger than the relative change seen for Ar$^+$, demonstrating that the initial alignment is a more critical factor for Ne$^+$ than
for Ar$^+$.

Figure \ref{fig:m1}(d) also shows that harmonics 17 - 29 have a very similar magnitude, apart from harmonic 25,
which has been reduced by over one order of magnitude, compared to the other harmonics in the
plateau region. It is noteworthy that, for $M=0$, both neighbouring harmonics, harmonics 23 and 27- were the ones that decreased noticeably in magnitude. To verify that the reason for this decrease is the interplay between channels associated with the $^3P^e$ and $^1D^e$ thresholds, as for $M=0$, we have carried out additional calculations in which combinations of these states were included as residual-ion states of Ne$^{2+}$.

\begin {table}[!hbtp]
\caption{Typical harmonic yields for Ne$^+$ with $M=1$ irradiated by 390 nm laser light at an intensity of 10$^{15}$
Wcm$^{-2}$ as a function of thresholds retained, normalized to the corresponding spectrum obtained with $M=0$. The cut-off energy of the plateau region and the final outer region population are also shown
for each subset of Ne$^{2+}$ thresholds retained.}

\begin{tabular*}{\columnwidth}{@{\extracolsep{\fill}}lccc}
\hline
\hline
Threshold  & Relative &            Cut-off           &     Population \\
           & harmonic yield &            \footnotesize{(eV)}                      &  in outer region  \\ 
           
           \hline
$^{1}D^{e}$,$^{3}P^{e}$ and $^{1}S^{e}$ & 26.4  & 99.2 & $1.05\times10^{-3}$ \\                         
$^{1}D^{e}$                             & 0.63       &	100.7 & $2.44\times10^{-4}$ \\ 
$^{3}P^{e}$                             &   	&	96.2 & $1.74\times10^{-3}$ \\
$^{1}D^{e}$,$^{3}P^{e}$         &18.2        &	99.2 & $1.15\times10^{-3}$ \\ 
\hline
\end{tabular*}
\label{tab:m=1}
\end{table}

Figure \ref{fig:m1}(a), \ref{fig:m1}(b) and \ref{fig:m1}(c) show the harmonic spectra obtained when
just the $^1D^e$ threshold is
retained, just the $^3P^e$ threshold is retained and when both the $^3P^e$ and $^1D^e$
thresholds are retained, respectively. For Figs. \ref{fig:m1}(a), \ref{fig:m1}(b) and \ref{fig:m1}(c), the
corresponding spectrum for
$M=0$ is also presented. Figure \ref{fig:m1}(c) and \ref{fig:m1}(d) demonstrate that
the total harmonic spectrum obtained
for the ($^3P^e$, $^1D^e$) case is very similar to the full spectrum, as observed for
$M=0$. The assumption that the $^1S^e$ threshold is less important is therefore justified. 

The comparison of the harmonic spectra for $M=1$ and $M=0$ is dramatically different for the
calculation including the $^3P^e$ threshold only and for the calculation including the $^1D^e$
threshold only. When only the $^1D^e$ threshold is included in the calculations, the harmonic
yields for $M=1$ and $M=0$ are very similar. The population in the outer region is also similar, differing by a factor 1.3. 
On the other hand, when only the $^3P^e$ threshold is retained, the harmonic intensities increase by
over 2 orders of magnitude. The intensities of the harmonics across the plateau region show
no obvious decrease with photon energy for $M=1$, whereas they did for $M=0$. 
This marked increase in harmonic yield is matched by an increase of a factor 12  in the population in the outer region.
This behaviour is in line with the behaviour seen for Ar$^+$ \cite{Bro4}. 

The spectrum obtained when both the $^3P^e$ and $^1D^e$ thresholds are retained in the
calculation shows harmonic intensities which have decreased from the typical intensity
shown in the $^3P^e$ spectrum by approximately a factor of 3. 
However, apart from harmonic 25, harmonics 17 - 29 appear with very similar intensity in
the ($^3P^e$, $^1D^e$) spectrum, whereas the $^3P^e$ spectrum shows more substantial variation
across the harmonics. Harmonic 25 shows a significant reduction of about an order of magnitude
compared to harmonics 17 - 29. In the individual $^3P^e$ and $^1D^e$ spectra, harmonic 25 is
similar in appearance as harmonic 23. Destructive interference between the two pathways to
HG leads to the significant reduction of the yield of harmonic 25 in the combined
spectrum.

The significant change seen in harmonic yield associated with the $^3P^e$ threshold has the same
origin as explained for Ar$^+$ \cite{Bro4}.
For $M=0$ the initial $2p^5$ configuration only contains a single electron with $m=0$. Its emission
leaves $2p^4$ in a singlet state, so the $^3P^e$ state can not be a state of the residual ion following ejection of an $m=0$ electron.
For $M=1$, two electrons have $m=0$.
Emission of one of these electrons leaves a $2p$ shell with
holes at $m=-1$ and $m=0$, which can combine to form a triplet state. Thus, the emission of
an $m=0$ electron can leave the Ne$^{2+}$ residual ion in the $^{3}P^e$ ground state for $M=1$. The ionization step
in the recollision model is dominated by a single $m=0$ electron escaping towards an excited
threshold for $M=0$, whereas it is dominated by one out $m=0$ electron (out of two available) escaping towards the
lowest threshold for $M=1$. Ionization should therefore be significantly stronger for $M=1$, and 
the harmonic yield should be higher. The $^1D^e$ threshold can be reached through emission of an $m=0$ electron for both $M=0$ and $M=1$. In this case, the harmonic yields and populations in the outer region are of similar magnitude.

In addition, Table \ref{tab:m=1}  also shows that  the cut-off energy, when only $^{3}P^e$ threshold is included in the $M=1$ calculation, is about 96 eV. On the other hand, it is about 99 eV when the $^1D^e$ threshold taken into account as well. The additional thresholds therefore appear to increase the cutoff energy for the harmonic plateau. 
For both $M=0$ and $M=1$, the inclusion of the $^1D^e$ threshold raises the cut-off energy. The role of the $^1D^e$ threshold differs in these cases: for $M=0$, it is the primary threshold for HG, but for $M=1$ it is the secondary threshold. 

Overall, the HG spectra for both $M=0$ and $M=1$ demonstrate the necessity to include at least the lowest two thresholds for the reliable determination of the HG spectrum, for the isolated atom. Interplay between channels associated with these thresholds affects the spectra greatly: the overall yield is reduced by a factor 4 from the most efficient channel. The cut-off energy is increased beyond the cut-off associated with the lowest threshold. Interference between pathways can cause specific harmonics to be significantly reduced at specific intensities.

\section{Conclusions}

We have applied time-dependent R-matrix theory to investigate HG in Ne$^{+}$ at
a laser wavelength of 390 nm. Due to the large binding energy of Ne$^{+}$, harmonic
spectra could be obtained for laser intensities up to $10^{15}$ Wcm$^{-2}$, enabling the
determination of the role of different ionization thresholds in HG within
the plateau region. To assess the influence of channels associated with individual thresholds and
of the interactions between channels associated with different ionization thresholds, calculations were
performed using all possible combinations of the three $1s^{2}2s^{2}2p^{4}$ $^3P^{e}$, $^1D^{e}$ and
$^{1}S^{e}$ thresholds of Ne$^{2+}$. A good approximation to the full spectrum is obtained when both
the $^{1}D^e$ and $^{3}P^e$ threshold are included, indicating that inclusion of the 
$^{1}S^{e}$ threshold is less critical.

For $M=0$, we find that inclusion of just the $^{3}P^e$ threshold on its own
underestimates the harmonic yield by up to an order of magnitude, and gives too small a cut-off energy. On the
other hand, inclusion of the $^{1}D^e$ threshold on its own overestimates the harmonic yield by an order of magnitude due to
an overestimation of the Ne$^+$ ionization rate. Hence both thresholds are essential for a correct description of
HG for Ne$^{+}$ with $M=0$.

For $M=1$, we find that inclusion of just the $^3P^e$ threshold on its own
slightly overestimates the harmonic yield by about a factor of 2, whereas, the cut-off energy value underestimates by 3 eV the value when all the three thresholds are retained. On the other hand, inclusion of only the $^{1}D^e$ threshold now underestimates the harmonic yield by an order of magnitude. Hence both thresholds are essential for a correct description of
HG for Ne$^{+}$ with $M=1$.

Interactions between the pathways leading up to the $^3P^e$ and $^1D^e$ thresholds affect more
than just the overall magnitude of the harmonics.
At an intensity of 10$^{15}$ Wcm$^{-2}$, we observe noticeable decreases in the harmonic yield for
specific harmonics. For $M=0$ harmonics 23 and 27 are affected in particular, whereas for $M=1$,
harmonic 25 is affected the most. These decreases in the harmonic yield are only
present when both the $^3P^e$ and the $^1D^e$ thresholds are included in the calculations, and
are not observed when only a single threshold is included. We thus ascribe this decrease to
interference between different ionization channels. It should be noted however that this
particular interference is only observed at 10$^{15}$ Wcm$^{-2}$, and is not observed at
9$\times$10$^{14}$ Wcm$^{-2}$. Therefore this interference
may be strongly intensity dependent. In experiment, where different atoms (or ions) will experience
different peak intensities, the effects of this interference may therefore not be apparent.

Although the present work demonstrates that the TDRM approach can describe HG in ions
successfully, including an extensive plateau up to photon
energies $\sim$100 eV, using intensities up to $10^{15}$ Wcm$^{-2}$ at a wavelength of 390 nm,
it may be difficult to extend the present approach to longer wavelengths due to an increase
in the number of angular momenta that need to be retained in the calculations.
The recently developed R-matrix
incorporating time-dependence (RMT) codes \cite{Moor2} should be more amenable to the
inclusion of many angular momenta. It will therefore be interesting to explore
HG using the RMT approach.

\section*{ACKNOWLEDGMENTS}
O. Hassouneh acknowledges the University of Jordan for financial support. 
H. W. H acknowledges financial support from UK EPSRC under Grant No. G/055416/1, and A. C. B support from DEL under the programme for government.

\end{document}